\documentclass[prl,showpacs,showkeys,nofootinbib,twocolumn,floatfix,superscriptaddress]{revtex4}

\usepackage{ulem}
\usepackage{graphicx}  %
\usepackage{amsmath}
\usepackage{bm}  %

\newcommand{\bea}{\begin{eqnarray}}
\newcommand{\eea}{\end{eqnarray}}
\newcommand{\beq}{\begin{equation}}
\newcommand{\eeq}{\end{equation}}
\newcommand{\bqa}{\begin{eqnarray}}
\newcommand{\eqa}{\end{eqnarray}}

\def\mqo2{{\!\!\!}}

\begin{document}

\title{
Three-body Recombination of $^6$Li Atoms with \\
Large Negative Scattering Lengths}

\author{Eric Braaten}
\affiliation{Department of Physics,
         The Ohio State University, Columbus, OH\ 43210, USA}

\author{H.-W. Hammer}
\affiliation{Helmholtz-Institut f\"ur Strahlen- und Kernphysik
        (Theorie) and Bethe Center for Theoretical Physics,
        Universit\"at Bonn, 53115 Bonn, Germany}
\author{Daekyoung Kang}
\affiliation{Department of Physics,
         The Ohio State University, Columbus, OH\ 43210, USA}

\author{Lucas Platter}
\affiliation{Department of Physics,
         The Ohio State University, Columbus, OH\ 43210, USA}
\date{\today}

\begin{abstract}
The 3-body recombination rate at threshold for 
distinguishable atoms with large negative pair scattering lengths
is calculated in the zero-range approximation.  
The only parameters in this limit are the 3 scattering lengths 
and the Efimov parameter, which can be complex valued. 
We provide semi-analytic expressions for the cases of 2 or 3 equal 
scattering lengths and we obtain numerical results for the general case 
of 3 different scattering lengths.
Our general result is applied to the three lowest hyperfine 
states of $^6$Li atoms.  Comparisons 
with recent experiments provide indications 
of loss features associated with Efimov trimers near the 3-atom threshold.
\end{abstract}

\smallskip
\maketitle

Atomic gases allow the experimental study of superfluidity 
in systems in which the fundamental interactions are simple 
and experimentally controllable.
In the case of fermionic atoms with two spin states,
there have been extensive investigations of the crossover 
from the {\it BCS mechanism} (Cooper pairing of atoms) 
to the {\it BEC mechanism} (Bose-Einstein condensation 
of diatomic molecules) \cite{KZ0801}.
Fermionic atoms with three spin states 
open up the possibility of new superfluid phases and
new mechanisms for superfluidity
\cite{Bedaque:2006ii,PMT06,HJZ06,Zhai06}.
The first experimental studies of such a system were
recently carried out using the three lowest hyperfine states 
of $^6$Li atoms \cite{Heidelberg,PennState}.

Fermionic atoms with three spin states also open up new possibilities 
in few-body physics. 
For the discussion of 3-body observables, they can simply be considered
as three distinguishable atoms and their fermionic nature plays no 
special role.
If the pair scattering lengths are large compared to the range
of the interactions between the atoms, a remarkable set of 
3-body phenomena are predicted.
If the 3 scattering lengths are infinitely large,
there is an infinite sequence of 3-atom bound states
called {\it Efimov trimers} with a geometric spectrum and an 
accumulation point at the 3-atom threshold \cite{Efimov}.
Low-energy 3-body phenomena governed by
{\it discrete scale invariance} are generally
referred to as {\it Efimov physics} \cite{Braaten:2004rn,Braaten-Hammer}.
The first experimental evidence 
for Efimov physics were experiments with ultracold 
$^{133}$Cs atoms by Grimm and 
coworkers in which they observed dramatic 
dependence of the 3-body recombination rate 
and the atom-dimer relaxation rate on the 
scattering length \cite{Grimm06,Grimm08}.

In this Letter, we present calculations of the
3-body recombination rate at threshold for 
distinguishable atoms with large negative pair scattering lengths
in the zero-range limit.
We provide semi-analytic expressions for the cases of 
2 or 3 equal scattering lengths 
and we obtain numerical results for the general case of 3 different 
scattering lengths.
We apply our general results to the three lowest hyperfine 
states of $^6$Li atoms and compare with recent 3-body recombination 
rate measurements \cite{Heidelberg,PennState}.

We consider an atom of mass $m$ with three distinguishable states
that we label 1, 2, and 3 and refer to as spin states.
We denote the scattering length of the pair $i$ and $j$
by either $a_{ij} = a_{ji}$ or $a_k$, where $(ijk)$
is a permutation of $(123)$.
The rate equations for the number densities $n_i$ 
of atoms in the three spin states are
\begin{equation}
\frac{d\ }{dt} n_i = - K_3 n_1 n_2 n_3.
\label{dndt}
\end{equation}
By the optical theorem, the event rate constant $K_3$  
in the low-temperature limit can be expressed 
as twice the imaginary part of the forward T-matrix element for 
3-atom elastic scattering in the limit where the momenta of the 
atoms all go to 0. 
Using diagrammatic methods, the T-matrix element for elastic scattering
can be expressed as the sum of 9 amplitudes corresponding to the 3
possible pairs that are the first to scatter and the 3 possible pairs
that are the last to scatter. For small collision energies, the leading
contributions to those amplitudes come from the S-wave terms, which we
denote by ${\cal A}_{ij} (p,p')$, where $p$ ($p'$) is the relative
momentum between the pair that scatters first (last) 
and the third atom labelled $i$ ($j$).
The rate constant $K_3$ in Eq.~(\ref{dndt}) is
\begin{equation}                               
K_3  = \frac{32 \pi^2}{m} \sum_{i,j} a_i a_j {\rm Im} {\cal A}_{ij}(0,0),
\label{K3-A}
\end{equation}
where the sums are over $i,j = 1,2,3$.
The amplitudes  ${\cal A}_{ij} (p,p')$ can be calculated in the
zero-range limit by solving 9 coupled integral equations that are
generalizations of the Skorniakov--Ter-Martirosian (STM) 
equation \cite{STM57}. 
To determine ${\rm Im} {\cal A}_{ij} (0,0)$, it is sufficient to 
solve the 9 coupled STM equations for ${\cal A}_{ij} (p,0)$:
\begin{eqnarray}
{\cal A}_{ij}(p,0) &=& \frac{1 - \delta_{ij}}{p^2}
+ \frac{2}{\pi} \sum_k (1 - \delta_{kj}) 
\nonumber
\\
&& \hspace{0.5cm} \times 
\int_0^\Lambda \! dq \, Q(q/p) D_k(q) {\cal A}_{ik}(q,0) ,
\label{STMeq}
\end{eqnarray}
where 
\begin{eqnarray}
Q(x) &=&  \frac{x}{2} \log \frac{1 + x + x^2}{1 - x + x^2},
\\
D_k(q)&=& (-1/a_k + \sqrt{3}q/2)^{-1},
\end{eqnarray}
and $\Lambda$ is an ultraviolet cutoff. 
The solutions to Eqs.~(\ref{STMeq})
are singular as $p \to 0$.  The singular terms, which  
are proportional to $1/p^2$, $1/p$, and $\ln p$,
appear only in ${\rm Re} \, {\cal A}_{ij}(p,0)$ for real $p$ 
and can be derived 
by iterating the integral equations \cite{Braaten:2001ay}.
Since ${\rm Im} \, {\cal A}_{ij}(p,0)$ must be extrapolated
to $p=0$, it is useful to transform Eqs.~(\ref{STMeq}) into
coupled STM equations for amplitudes $\bar {\cal A}_{ij}(p,0)$ obtained 
by subtracting the singular terms from ${\cal A}_{ij}(p,0)$.
For $p \ll \Lambda$,
the solutions depend log-periodically on $\Lambda$ with a discrete
scaling factor $e^{\pi/s_0} \approx 22.7$, where $s_0 = 1.00624$.
The dependence on the arbitrary cutoff $\Lambda$ can be 
eliminated in favor of a physical 3-body parameter, such as
the Efimov parameter $\kappa_*$ defined by the spectrum of Efimov states 
in the limit where all 3 scattering lengths are infinitely large 
\cite{Braaten:2004rn}:
\begin{equation}                               
E_n = - \left( e^{2 \pi/s_0} \right)^{-n} \frac{\hbar^2 \kappa^2_*}{m}
 ~~~ (a_{12} = a_{23} = a_{31} = \pm \infty)  .
\label{Efimov}
\end{equation}
If we restrict $\Lambda$ to a range that
corresponds to a multiplicative factor of 22.7, then $\Lambda$ 
differs from $\kappa_*$ only by a multiplicative numerical
constant. Thus we can also simply take $\Lambda$ as the 3-body
parameter.

If $a_{ij} > 0$, there is a contribution to $K_3$ from
3-body recombination into the shallow dimer whose constituents 
have spins $i$ and $j$ and whose binding energy is 
$\hbar^2/(m a_{ij}^2)$.  If $a_{12}$, $a_{23}$, and $a_{31}$ 
are all negative, there are no shallow dimers.
The solutions ${\cal A}_{ij} (p,0)$ to the coupled STM
equations in Eq.~(\ref{STMeq}) are all real-valued in this case, 
so the rate constant $K_3$ in Eq.~(\ref{K3-A}) is predicted to be 0. 

If there are deeply-bound diatomic molecules ({\it deep dimers})
in any of the three 2-body
channels, there are also contributions to $K_3$ 
from 3-body recombination into the deep dimers.
If all 3 scattering lengths are negative,
these are the only contributions to $K_3$.
The coupled STM equations in Eq.~(\ref{STMeq}) do not 
take into account contributions from deep dimers.
The inclusive effect of all the deep dimers 
can be taken into account by analytically continuing 
the Efimov parameter $\kappa_*$ to a complex value \cite{Braaten:2003yc}: 
$\kappa_* \to \kappa_* \exp (i \eta_* / s_0)$, where $\eta_*$ is a
positive real parameter. 
Making this substitution in Eq.~(\ref{Efimov}), 
we find that the Efimov states acquire nonzero decay widths 
determined by $\eta_*$.  If we use the ultraviolet cutoff $\Lambda$ 
as the 3-body parameter, the inclusive effects of deep dimers 
can be taken into account by changing the upper limit of the integral 
in Eq.~(\ref{STMeq}) to $\Lambda \exp (i \eta_* / s_0)$,
so the path of integration extends into the complex plane. 
Having made this change, the solutions ${\cal A}_{ij} (p,0)$
are complex-valued even if $a_{12}$, $a_{23}$ and $a_{31}$ are all
negative. The rate constant $K_3$ in Eq.~(2) is a 
function of the scattering lengths $a_{12}$, $a_{23}$, and $a_{31}$
and the 3-body parameters $\Lambda$ and $\eta_*$ and it
vanishes as $\eta_* \to 0$. It gives the inclusive rate 
for 3-body recombination into all deep dimers.

\begin{figure}[t]
\centerline{\includegraphics*[height=5cm,angle=0,clip=true]{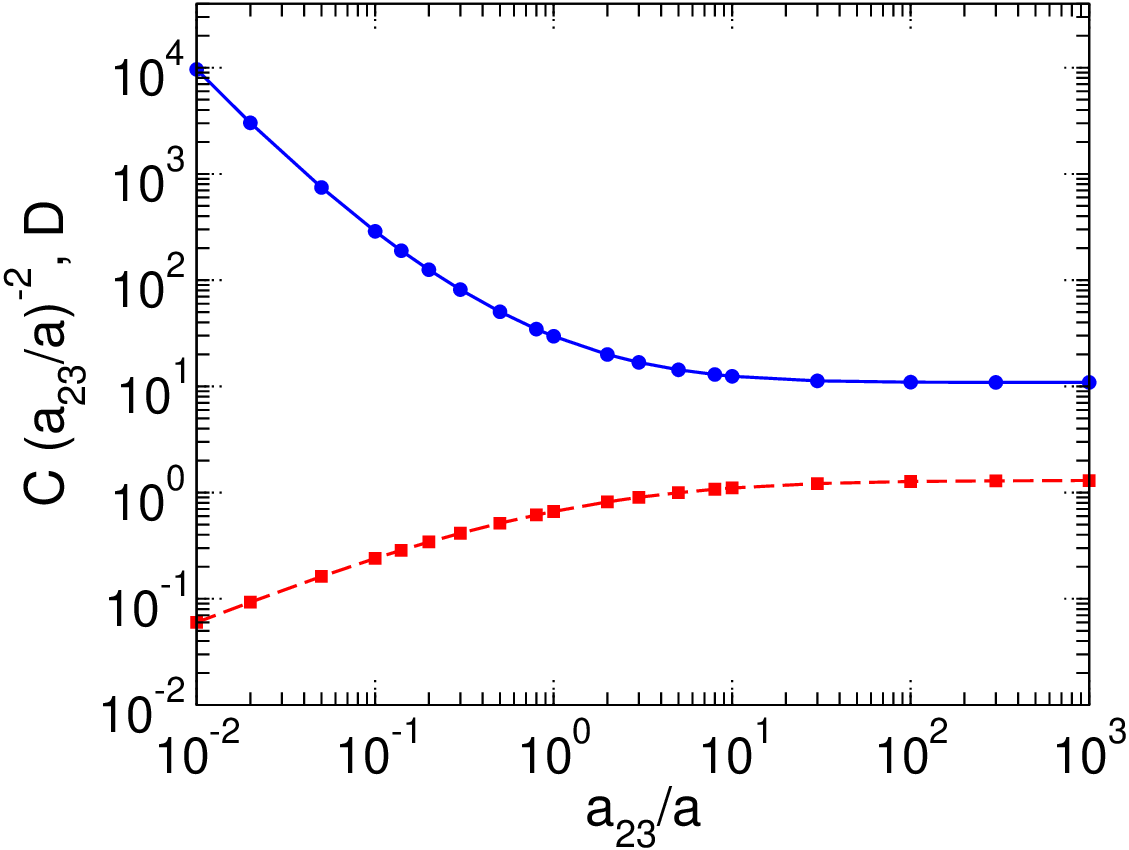}}
\vspace*{0.0cm}
\caption{(Color online) 
The coefficients $C$ scaled by $(a_{23}/a)^{-2}$ 
(upper curve) and $D$ (lower curve) 
in Eq.~(\ref{K3-equal}) as functions of $a_{23}/a$ 
for the case of two equal negative scattering lengths
$a$ and a third negative  scattering length $a_{23}$.
}
\label{fig:2a}
\end{figure}

We focus our attention on cases in which all
scattering lengths are negative, so the only 
recombination channels are into deep dimers.
We first consider the case of 3 equal 
scattering lengths: $a_{12} = a_{23} = a_{13} = a < 0$.
In this case Eq.~(\ref{STMeq}) reduces -- after summing
over $i$ and $j$ -- to the STM equation for identical bosons.
In Ref.~\cite{Braaten:2003yc}, Braaten and Hammer deduced an 
analytic expression for the 3-body recombination rate constant 
for identical bosons with a large negative scattering length $a$:
\beq
K_3 =
\frac{16 \pi^2 C \sinh(2 \eta_*)}
    {\sin^2[s_0 \ln(D |a| \kappa_*)] + \sinh^2 \eta_*}
\, \frac{\hbar a^4}{m}~,
\label{K3-equal}
\eeq
where $s_0 = 1.00624$, $C$, 
and $D$ are numerical constants.
This formula exhibits resonant enhancement for $a$ near 
the values $(e^{\pi/s_0})^n (D \kappa_*)^{-1}$ 
for which there is an Efimov state at the 3-body threshold.
Fitting our numerical results for $K_3/a^4$ as functions of 
$a \Lambda$ and $\eta_*$, we determine the numerical constants 
to be $C = 29.62(1)$ and  $D = 0.6642(2)$.
These values are more accurate than previous results for
identical bosons \cite{Braaten:2004rn}.
A separate calculation of the spectrum 
of Efimov states in the limit $a \to \pm \infty$ with $\eta_* = 0$
is necessary to determine the relation between the Efimov parameter 
and the ultraviolet cutoff: $\kappa_* = 0.17609(5) \Lambda$.

We next consider the case of 2 equal negative scattering 
lengths and a third that vanishes:
$a_{12} = a_{13} = a< 0$, $a_{23}=0$.
In this case with only two resonant scattering channels, 
$s_0 = 0.413698$ and
the discrete scaling factor is $e^{\pi/s_0} \approx 1986$.
Eq.~(\ref{K3-equal}) again gives an excellent fit 
to our numerical results and we determine the 
numerical constants as $C = 0.8410(6)$ and  $D = 0.3169(1)$.

We now consider the case of 2 equal negative scattering 
lengths and a third that is unequal:
$a_{12} = a_{13} = a < 0$, $a_{23}< 0$.
Eq.~(\ref{K3-equal}) with $s_0 = 1.00624$ continues 
to provide an excellent fit to our numerical results.
The fitted values of $C$ and $D$
are shown as functions of $a_{23}/a$ in Fig.~\ref{fig:2a}.
For $|a_{23}| \gg |a|$, the coefficients seem to have the 
limiting behaviors $C \approx 10.88(2)~(a_{23}/a)^2$ and $D \approx 1.30(1)$.
Their limiting behaviors for $|a_{23}| \ll |a|$ do not seem to be 
simple power laws. 
This is not surprising, because the discrete 
scaling factor 22.7 changes to 1986 when $a_{23} = 0$.

\begin{figure}[t]
\centerline{\includegraphics*[height=3.8cm,angle=0,clip=true]{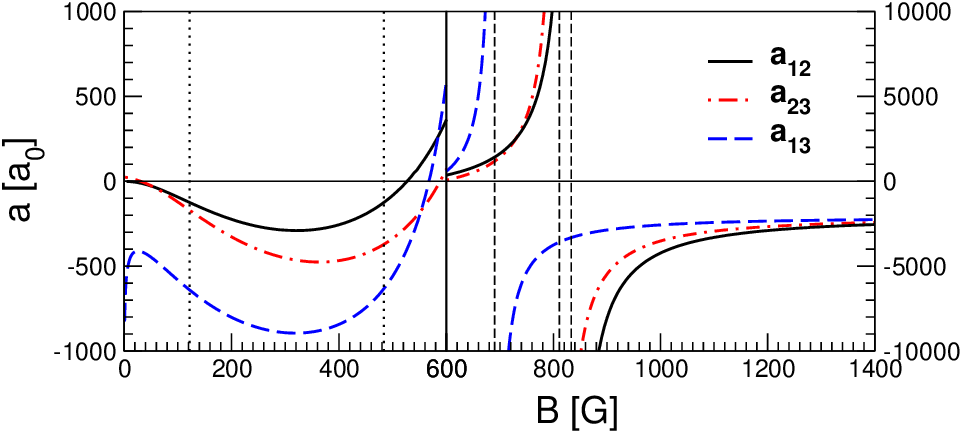}}
\vspace*{0.0cm}
\caption{(Color online) 
The scattering lengths for the three lowest hyperfine states 
of $^6$Li as functions of the magnetic field $B$ \cite{Julienne}.
The vertical scale changes by a factor of 10 at $B = 600$~G.
The two vertical dotted lines mark the boundaries of a region
in which $|a_{12}| > 2~\ell_{\rm vdW}$.
The three vertical dashed lines mark the positions of the 
Feshbach resonances. 
}
\label{fig:scatteringlengths}
\end{figure}

Finally we consider the general problem of 3 different 
negative scattering lengths, 
for which we can obtain numerical results
for given values of $a_{12}$, $a_{23}$, and $a_{13}$. 
We apply our method to $^6$Li atoms 
in the three lowest hyperfine states $|f, m_f \rangle$:
$| 1 \rangle = |\frac12, +\frac12 \rangle$,
$| 2 \rangle = |\frac12, -\frac12 \rangle$, and
$| 3 \rangle = |\frac32, -\frac32 \rangle$.
The 3 pair scattering lengths $a_{12}$, $a_{23}$, and $a_{13}$
are shown as functions of the magnetic field in 
Fig.~\ref{fig:scatteringlengths} \cite{Julienne}.
They have Feshbach resonances near 834~G, 811~G, and 690~G, 
respectively \cite{6Li}.  The zero-range approximation 
should be accurate if $|a_{12}|$, $|a_{23}|$, and $|a_{13}|$
are all much larger than the van der Waals length 
$\ell_{\rm vdW} = (m C_6/\hbar^2)^{1/4}$, 
which is approximately $62.5~a_0$ for $^6$Li.
There are two regions of the magnetic field in which all 
3 scattering lengths are negative and satisfy 
$| a_{ij}| > 2 \ell_{\rm vdW}$:
a low-field region $122~{\rm G} < B < 485~{\rm G}$
and a high-field region $B > 834~{\rm G}$.
In the low-field region, the smallest scattering length 
is $a_{12}$ and it achieves its largest value 
$- 290~a_0 = -4.6~\ell_{\rm vdW}$ near 320~G.
The zero-range approximation may
be reasonable near this value of $B$.
In the high-field region, 
the smallest scattering length is $a_{13}$.
It increases from $-3285~a_0$ at $B=834$~G 
to $-2328~a_0 \approx -37~\ell_{\rm vdW}$ at 1200~G. 
Thus the zero-range approximation should be very accurate
in this region.
We emphasize that the 3-body parameters
$\kappa_*$ and $\eta_*$ need not be the same in the two
universal regions, since there are
zeroes of the scattering lengths between them.

\begin{figure}[t]
\centerline{\includegraphics*[height=5.0cm,angle=0,clip=true]{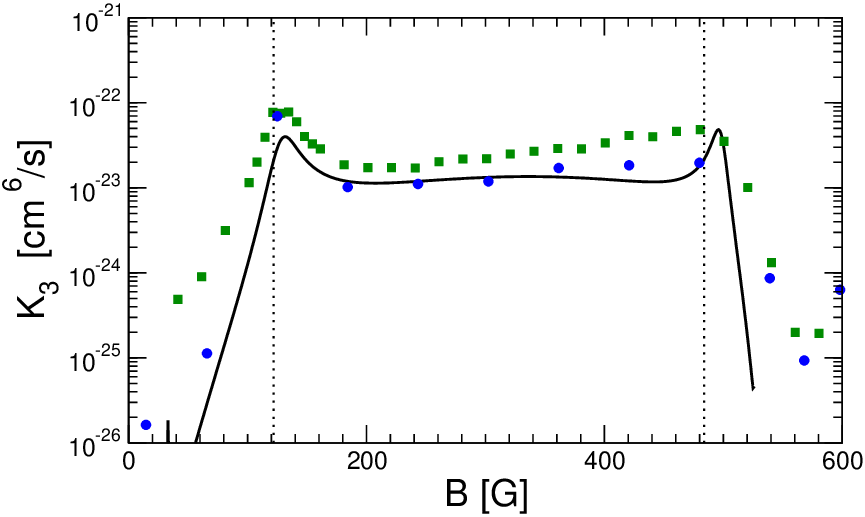}}
\vspace*{0.0cm}
\caption{(Color online) 
The 3-body recombination rate constant $K_3$ 
as a function of the magnetic field $B$.
The two vertical dotted lines mark the boundaries of the region
in which $|a_{12}| > 2~\ell_{\rm vdW}$.
The solid squares and dots are data points from Refs.~\cite{Heidelberg}
and \cite{PennState}, respectively.
The curve is a 2-parameter fit to the shape of the 
data from Ref.~\cite{Heidelberg}.}
\label{fig:Klow}
\end{figure}

\begin{figure}[t]
\centerline{\includegraphics*[height=5.0cm,angle=0,clip=true]{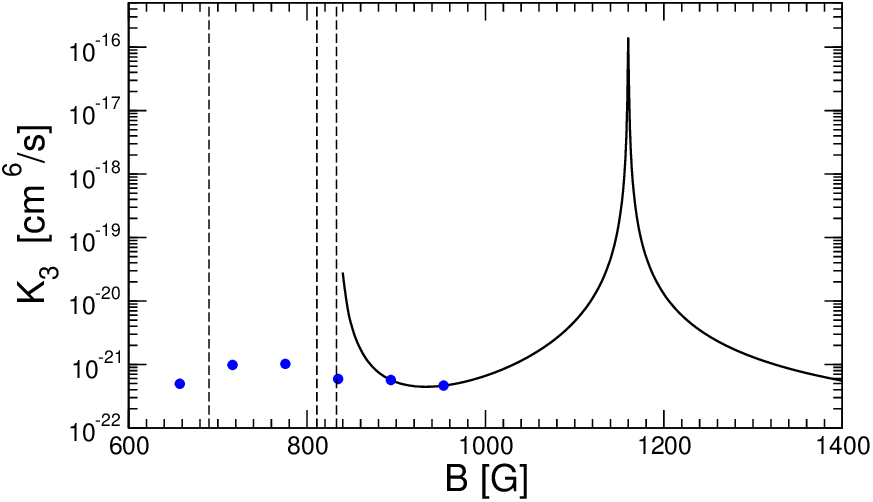}}
\vspace*{0.0cm}
\caption{(Color online) 
The 3-body recombination rate constant $K_3$ 
as a function of the magnetic field $B$.
The three vertical dashed lines mark the positions of the 
Feshbach resonances.
The solid dots are data points from Ref.~\cite{PennState}.
The curve is a 2-parameter fit to the last two data points.}
\label{fig:Khigh}
\end{figure}

The 3-body recombination rate $K_3$ for $^6$Li atoms in the three
lowest hyperfine states has recently been measured by
Jochim et al.\ \cite{Heidelberg}
and by O'Hara et al.\ \cite{PennState}.
Their results are shown in Figs.~\ref{fig:Klow} and \ref{fig:Khigh}.
In Ref.~\cite{Heidelberg}, $K_3$ was measured for each of the 
three spin states separately.
Those results have been averaged to get a single value of $K_3$ 
at each value of $B$.
Both groups observed dramatic variations in $K_3$ with $B$,
including a narrow loss feature near 130~G 
and a broader loss feature near 500~G.

The narrow loss feature and the broad loss feature observed in
Refs.~\cite{Heidelberg,PennState} appear near the 
boundaries of the low-field region in which all 3 scattering 
lengths satisfy
$|a_{ij}| > 2 \ell_{\rm vdW}$.  The zero-range approximation 
is questionable near the boundaries of this region.
We nevertheless fit the data for $K_3$ in this region 
by calculating the 3-body recombination rate using the 
$B$-dependence of $a_{12}$, $a_{23}$, and $a_{13}$ 
shown in Fig.~\ref{fig:scatteringlengths}, while
treating $\Lambda$ and $\eta_*$ as fitting parameters.
Since the systematic error in the normalization of $K_3$ 
was estimated to be $90\%$ in Ref.~\cite{Heidelberg}
and $70\%$ in Ref.~\cite{PennState}, we 
only fit the shape of the data and not its normalization.
A 2-parameter fit to the data from Ref.~\cite{Heidelberg} 
in the region $122~{\rm G} < B < 485~{\rm G}$ gives 
$\Lambda = 436~a_0^{-1}$ and $\eta_* = 0.11$.
The fit to the shape of the narrow loss feature
is excellent as shown in Fig.~\ref{fig:Klow}.
Having fit the position and width of the loss feature 
feature, the normalization of 
$K_3$ is determined.  
In the region of the narrow loss feature, the prediction for $K_3$ 
lies below the data of Ref.~\cite{Heidelberg} by about a factor of 2, 
which is well within the systematic error of $90\%$.
The excellent fit to the shape of the narrow loss feature 
and the prediction of the normalization of $K_3$ 
consistent with the data suggests that this loss feature 
may arise from an Efimov state near the threshold 
for atoms in spin states 1, 2, and 3.
As shown in Fig.~\ref{fig:Klow}, our fit predicts that $K_3$
should be almost constant in the middle of the low-field region
and that there should be another narrow loss feature
at its upper end near 500~G.
The data from both groups in Fig.~\ref {fig:Klow} increases monotonically
in the middle of the low-field region and, instead of a
narrow loss feature, there is a broad loss feature
near the upper end of this region.
We are unable to get a good fit to the slope of $\log K_3$
in the middle of the low-field region or to the shape of the
broad loss feature by adjusting
$\Lambda$ and $\eta_*$.

In Ref.~\cite{PennState}, the 3-body recombination rate 
was also measured at higher values of the magnetic field.
They include three data points in the region $B>834$~G,
where all 3 scattering lengths are extremely large 
and negative.
If the central values of the last two data points 
are used to determine the 3-body parameters,
we obtain $\Lambda = 37.0~a_0^{-1}$ 
and $\eta_* = 2.9 \times 10^{-4}$.
As shown in Fig.~\ref{fig:Khigh}, this fit predicts the 
resonant enhancement of the 3-body recombination rate near 1160~G.
If we allow for the systematic error by increasing 
or decreasing both data points by 70\%, the position of the 
resonance does not change, but $\eta_*$ increases to
$5 \times 10^{-4}$ or decreases to $9 \times 10^{-5}$, 
respectively.  If we take into account the statistical errors 
by increasing or decreasing the data points 
by one standard deviation, the position of the 
resonance can be shifted downward to 1109~G or upward to 1252~G. 
Thus it might be worthwhile to search for an Efimov resonance 
in this region. If such a feature were observed, measurements of 
its position and width would determine 
accurately the two 3-body parameters $\kappa_*$ and $\eta_*$.
Our equations could then be used to predict the total
3-body recombination rate in the entire universal region 
$B>610$~G, including the regions where 1, 2, or 3 of the 
scattering lengths are positive.
Note that the third-to-last data point in Fig.~\ref{fig:Khigh}
shows no sign of the large increase in $K_3$ near the Feshbach 
resonance at 834~G that is predicted by our fit.
However the measurement of $K_3$ involves a model for the 
heating of the system, and the failure of our fit at 835~G
might be attributable to the breakdown of that model 
near the Feshbach resonance.

In summary, we have calculated the recombination rate of three distinguishable
atoms with large negative pair scattering lengths in the zero-range limit.
We have provided simple semi-analytical expressions for the rate 
if 2 or 3 scattering lengths are equal. 
Using our general result for 3 unequal scattering lengths, we showed
that the narrow 3-body loss feature for $^6$Li atoms with three spin states
\cite{Heidelberg,PennState} may be attributed
to an Efimov state near threshold.
Our results provide a starting point for
a quantitative understanding of the 3-body loss rates and the unambiguous
identification of Efimov physics in these systems.

\begin{acknowledgments}
We thank S.~Jochim, K.M.~O'Hara, and A.N.~Wenz for useful communications.
This research was supported in part by the 
DOE under grants DE-FG02-05ER15715 and DE-FC02-07ER41457,
by the NSF under grant PHY-0653312, and by the BMBF under contract 06BN411.
\end{acknowledgments}

\end{document}